\documentclass[]{aa} 
\usepackage{graphicx}
\usepackage{txfonts}
\usepackage{natbib}
\bibpunct{(}{)}{;}{a}{}{,}
\begin{document}
   \title{On the metallicity gradient of the Galactic disk\thanks{Based on 
    observations made with ESO Telescopes in La Silla Observatory under 
    the program: 60.A-9120(B)}}


   \author{S.\ Pedicelli\inst{1,2,3} 
          \and G.\ Bono\inst{1,3}
	  \and B.\ Lemasle\inst{4}
          \and P.\ Fran\c{c}ois \inst{5}
          \and M.\ Groenewegen \inst{6}
          \and J.\ Lub,\inst{7}
          \and J.\ W.\ Pel,\inst{8} 
          \and D.\ Laney,\inst{9}
          \and A.\ Piersimoni,\inst{10}
          \and M.\ Romaniello \inst{2}
          \and R.\ Buonanno,\inst{1}
          \and F.\ Caputo,\inst{3}
          \and S.\ Cassisi,\inst{10}
          \and F.\ Castelli,\inst{11}
          \and S.\ Leurini,\inst{2}
          \and A.\ Pietrinferni,\inst{10}
          \and F.\ Primas \inst{2}
          \and J.\ Pritchard\inst{2}
          }
   \institute{Universit\`a di Roma Tor Vergata, Via della Ricerca Scientifica 1, 00133 Roma, Italy; \email{pedicelli@mporzio.astro.it}
    \and European Southern Observatory (ESO) Karl-Schwarzschild-Strasse 2, D-85748 Garching bei M{\"{u}}nchen, Germany 
    \and INAF -- Osservatorio Astronomico di Roma, Via Frascati 33, Monte Porzio Catone, Italy
    \and Universit\'e de Picardie Jules Verne, Facult\'e des Sciences, 33 rue Saint-Leu, 80039 Amiens Cedex 1, France
    \and Observatoire de Paris-Meudon, GEPI, 61 avenue de l'Observatoire, F-75014 Paris, France
    \and Royal Observatory of Belgium, Ringlaan 3  B-1180 Brussels, Belgium
    \and Leiden Observatory, Leiden University, P.O. Box 9513, NL-2300 RA  Leiden, The Netherlands
    \and Kapteyn Institute, University of Groningen, P.O. Box 800, 9700 AV Groningen, The Netherlands
    \and South African Astronomical Observatory, PO Box 9, 7935 Observatory, South Africa
    \and INAF -- Osservatorio Astronomico di Collurania, Via M. Maggini, 64100 Teramo, Italy
    \and INAF -- Osservatorio Astronomico di Trieste, Via G.B. Tiepolo 11, 34143 Trieste, Italy 
}

   \date{}

 
  \abstract
    {}
   {The iron abundance gradient in the Galactic stellar disk provides 
    fundamental constraints on the chemical evolution of this important 
    Galaxy component. However the spread around the mean slope is, at 
    fixed Galactocentric distance, larger than estimated uncertainties. 
   }
   {To provide quantitative constraints on these trends we adopted iron abundances 
    for 265 classical Cepheids (more than 50\% of the currently known sample) based 
    either on high-resolution spectra or on photometric metallicity indices. Homogeneous  
    distances were estimated using near-infrared Period-Luminosity relations. The sample 
    covers the four disk quadrants and their Galactocentric distances range from  
    $\sim$5 to $\sim$17 kpc. We provided a new theoretical calibration of 
    metallicity-index-color (MIC) relation based on Walraven and NIR photometric 
    passbands. 
    We estimated the photometric metallicity of 124 Cepheids. Among them 66 
    Cepheids also have spectroscopic iron abundances and we found that the 
    mean difference is $-0.03\pm0.15$ dex. We also provide new iron abundances,  
    based on high-resolution spectra, for four metal-rich Cepheids located in 
    the inner disk. The remaining iron abundances are based on high-resolution 
    spectra collected by our group (73) or available in the literature (130).
   }
   {A linear regression over the entire sample provides an iron gradient of  
    -0.051 $\pm$ 0.004 dex $kpc^{-1}$. The above slope agrees quite well, within 
    the errors, with previous estimates based either on Cepheids or on open clusters 
    covering similar Galactocentric distances. However, Cepheids located in the 
    inner disk appear systematically more metal-rich than the mean metallicity gradient. 
    Once we split the sample in inner ($R_G <$8 kpc) and outer disk Cepheids we found 
    that the slope (-0.130$\pm$0.015 dex $kpc^{-1}$) in the former region is $\approx$3 
    times steeper than the slope in the latter one (-0.042 $\pm$ 0.004 dex $kpc^{-1}$). 
    We found that in the outer disk the radial distribution of metal-poor 
    (MP, $[Fe/H]<-0.02$ dex) and metal-rich (MR) Cepheids across the four disk quadrants 
    does not show a clear trend when moving from the innermost to the external disk
    regions. We also found that the relative fractions of MP and MR Cepheids in the 
    1st and in the 3rd quadrant differ at 8$\sigma$ (MP) and 15$\sigma$ (MR) level.  
    Finally, we found that iron abundances in two local overdensities of the 
    2nd and of the 4th quadrant cover individually a range in iron abundance 
    of $\approx$0.5 dex.      
   } 
   {Current findings indicate that the recent chemical enrichment across the 
    Galactic disk shows a clumpy distribution.}
   

   \maketitle
%

\section{Introduction}
Abundance gradients across the Galactic stellar disk 
provide fundamental constraints on the chemical evolution of this 
Galaxy component and on the plausibility of the physical assumptions 
adopted in chemical evolution models \citep{and04,lu06,le08}.  
Although the abundance gradients have been the cross-road of 
several empirical \citep{fri02, carr07} and theoretical 
\citep{por99, por00, chiap01, ces07} investigations, we still lack quantitative 
constraints on the observed spread in chemical composition at fixed 
Galactocentric distance \citep{lu06, le07}. 
Moreover, we need to assess whether different stellar tracers give similar 
trends when moving from the inner to the outer disk regions.    
Galactic Cepheids, when compared with other tracers (open clusters, 
$HII$ regions, $B$-type stars, Planetary Nebulae), 
present several advantages to evaluate elemental abundances in the Galactic disk: 
{\em i)}-- they are luminous and easily identified objects; {\em ii)}-- they are 
ubiquitous across the disk; {\em iii)}-- their spectra show a wealth of well 
defined lines, therefore, accurate abundance measurements of iron and heavy 
elements can be provided. The more recent estimates of iron abundance gradients 
based on Cepheids, provide slopes ranging from $\sim$-0.05 dex 
$kpc^{-1}$ \citep{cap01, and02c, lu03, kov05, lu06, yo06, le07} 
to $\sim$-0.07 dex $kpc^{-1}$ \citep{le08}. 
By using a sample of 40 open clusters located between the solar circle and 
$R_G \sim 14$ kpc \citet{fri02} found a slope of -0.06 dex $kpc^{-1}$. More 
recently, \citet{carr07} using new accurate metal abundances for five old open 
clusters located in the outer disk ($12 \le R_G \le 21$ kpc) and the sample adopted 
by \citet{fri02} found a much shallower global iron gradient, namely 
-0.018 dex $kpc^{-1}$. 
By using oxygen abundances of $HII$ regions, with Galactocentric distances 
ranging from 5 to 15 kpc, \citet{deh00} also suggest a mild 
slope (-0.04 dex $kpc^{-1}$), but these young objects do not show evidence of 
a flattening of the gradient in the outer disk.
Moreover, a sharp change in the metallicity gradient for $R_G \sim 10 - 12$ kpc 
was suggested by \citet{twa97} using 76 open cluster with distances ranging from 
6 to 15 kpc. In particular, they found that a good fit of the metallicity gradient 
can be provided using two zones, namely an inner disk with $6 \le R_G \le 10$ kpc 
and an outer disk with $R_G > 10$ kpc. The two zones are characterized by 
shallow slopes and by a discontinuity of $\approx$-0.2 dex at $R_G\approx 10$ kpc.  
This hypothesis was supported by \citet{and02c}, \citet{lu03}, 
\citet{and04} and by \citet{cap01}. However, these estimates might be hampered 
by the limited number of tracers close to the edge of the inner disk 
($R_G \sim 3-5$ kpc) and in the outskirts of the outer disk \citep{le08}.  
In this paper we investigate the Galactic metallicity gradient using a large 
sample of both photometric and spectroscopic Cepheid abundances. 


\section{Photometric and spectroscopic data} 
Multiband ($V$, $B$, $L$, $U$, $W$) Walraven photometry for 173 Galactic Cepheids was 
collected in several observing runs 1962 \citep{w64} and 1970-1971 \citep{p76,p78}  
at the Leiden Southern Station (South Africa).  The sample is 82\% complete 
for all known Cepheids brighter than $V=11.0$ mag at minimum light and south 
of declination +15$^{\circ}$ \citep{f95}. For each object were collected at 
least 30 phase points that properly cover the entire pulsation cycle. This 
means that the 
intrinsic accuracy of the mean magnitudes determined by fitting a cubic 
spline is better than a few hundredths of magnitude.
Note that these data were transformed into the standard Walraven system using 
data collected in La Silla \citep[1979-1991,][]{pelub07}.  
Together with optical photometry we also collected for the same objects accurate 
multiband $J,H,K$ Near-Infra-Red (NIR) data. These data are available for 
98 Cepheids of the Walraven sample. Among them, 92 were collected at 
SAAO (Laney \& Stobie 1994, [38 objects]; van Leeuwen et al.\ 2007, [45 objects]; 
unpublished [9 objects]). 
The uncertainty of each phase point ranges from $0.005$ to $0.007$ for $K < 6$ mag, 
deteriorating to about $0.012$ at $K=8.6$ mag. This implies an accuracy in the mean 
magnitudes of $\sim 0.002-0.005$ mag, depending on the number of points. 
However, the dominant source of uncertainty in the mean magnitudes is due to the 
absolute calibration and it is $\sim 0.01$ mag. Multiband NIR photometry for six 
Cepheids (V336 Aql, V600 Aql, RZ CMa, DX Gem, AY Sgr and CK Sct, Pedicelli et al.\ 2009 
in preparation) were collected with the 1.1m telescope AZT-24 available in Campo 
Imperatore (Aquila). For each object we secured $\approx$ a dozen of random observations 
per band and the typical uncertainty on the mean magnitudes is $\sim 0.02$ mag. 
Note that the difference in the accuracy between optical and NIR mean magnitudes 
is mainly due to the difference in the luminosity amplitude. The amplitudes in the 
former passbands are on average $\sim$2--3 times larger than in the latter ones.    
The mean optical and NIR magnitudes were determinated as a time-average along the 
linear intensity light curve and then transformed into magnitude. The mean colors 
were determined as the difference of the mean magnitudes involved in the color index.   
The mean NIR magnitudes of the other Cepheids were estimated using the 2MASS catalog 
\citep{cu03} and the template light curve from \citet{sos05}. The optical amplitudes 
and the epoch of maximum were retrieved from the Fernie 
catalog\footnote{http://www.astro.utoronto.ca/DDO/research/cepheids/} \citep{f95} 
and from the McMaster catalog\footnote{http://crocus.physics.mcmaster.ca/Cepheid/}. 
We compared individual reddening values provided by \citet{f95} and by 
\citet{lancad07} and for the objects in common we found differences of 
the order of a few percent. To overcome possible systematic uncertainties 
in the metallicity estimate, we adopted for the entire sample the reddenings 
provided by \citet{f95}. Selective absorptions in the Walraven bands 
were estimated following \citet{ped08}: $E(B-V)_J/E(V-B)=2.375-0.169 \cdot (V-B)$, 
$A_{V}/E(V-B)=3.17-0.16 \cdot (V-B)-0.12 \cdot E(V-B)$ and $E(B-L)/E(V-B)=0.39$ mag. 
Individual distances were estimated using the $J$,$H$,$K$ mean magnitudes 
and the empirical NIR Period-Luminosity (PL) relations for fundamental 
(FU) Cepheids provided by \citet{per04}. To estimate the distance of the 
eight first overtone (FO) Cepheids (FF Aql, GH Car, AZ Cen, BB Cen, V419 Cen, 
EV Sct, AH Vel, BG Vel) in our sample their periods were fundamentalized, 
i.e. we added 0.127 to their logarithmic period \citep{pl78, f95, le08}. 
Note that the use of more recent NIR PL relations \citep{fou07, vanl07, groe08} 
does not affect the conclusions of this investigation (see the error bars in 
the top panel of Fig.\ref{Fig3}).  
Selective absorptions in the $NIR$-bands were estimated using 
the reddening law by \citet{card89}. We adopted an LMC true distance modulus 
of 18.50 mag \citep{free01} and the heliocentric distances were estimated as 
the mean of the three distances in $J$, $H$, and $K-$band. The typical 
standard deviation for the mean distance modulus is smaller than 0.1 mag.   

\begin{figure}
   \centering
   \includegraphics[width=7cm]{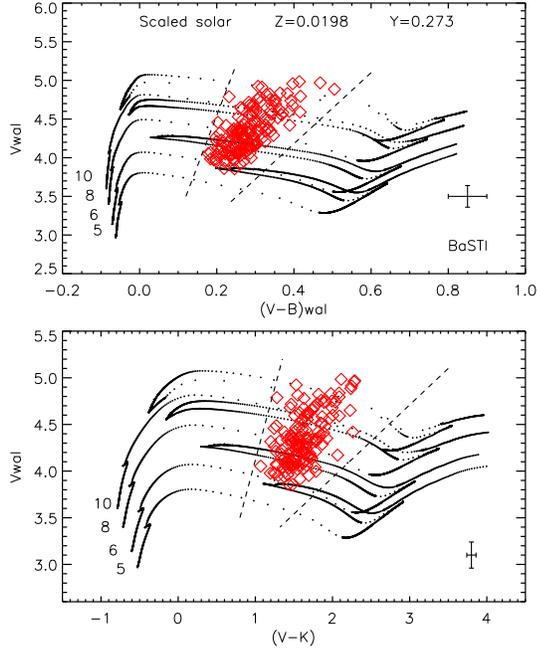}
   \caption{Top: Walraven $V$,$B-V$ Color-Magnitude Diagram (CMD) for Galactic  
Cepheids (red diamonds). Solid lines display evolutionary tracks at solar 
chemical composition (Z=0.0198, Y=0.273) and different stellar masses 
(see labeled values) from the BaSTI data set. The dashed lines show the edges 
of the instability strip for solar chemical composition. 
Individual Cepheid distances were estimated using the NIR PL relations by 
\citet{per04} and reddening estimates provided by \citet{f95}. The error 
bars account for uncertainties on the mean reddening correction (5\%) 
and on distances. Bottom: same as the top, but in the  $V$,$V-K$ CMD.}
              \label{Fig1}
    \end{figure}

In order to validate current estimates of reddenings and distance moduli 
we compared theory and observations using scaled solar evolutionary tracks 
from the BaSTI database\footnote{http://albione.oa-teramo.inaf.it/index.html} 
\citep{pi04, pi06}. Therefore, theoretical predictions were   
transformed into the Walraven bands using the bolometric corrections and 
the color-temperature relations provided by \citet{cast03}.  
Fig. \ref{Fig1} shows the comparison between theory and observations for 
the Walraven $V$,$B$-bands (top panel) and for optical-NIR $V$,$K$-bands 
(bottom panel). 
The error bars in the right corner account for uncertainties on the mean 
reddening correction (5\%) and on the absolute distance (sum in quadrature  
of the intrinsic dispersions of the NIR PL relations provided by \citet{per04}
and on the mean NIR meagnitudes). The solid lines display evolutionary tracks 
at solar chemical composition (metals, Z=0.0198; helium, Y=0.273) and mass values 
ranging from 5 to 10 $M_\odot$. The dashed lines display the predicted 
FO blue edge (hotter) and the FU red edge (cooler) 
of the Cepheid instability strip \citep{bo05} at solar chemical composition. 
Data plotted in this figure indicate that evolutionary and pulsation 
predictions agree quite well with observations. This outcome applies to 
both faint (short period) and bright (long period) Cepheids.  

The spectroscopic data set includes: ten Cepheids from \citet{ro08}, 
63 from \citet{le08,le07}, six from \citet{szi07}, eight from \citet{lu06} 
and 116 from \citet[][b, c]{and02a}. These data were complemented with accurate 
iron abundances for four inner disk Cepheids (AV Sgr; V340 Ara; VY Sgr; UZ Sct). 
A more detailed discussion concerning photometric and spectroscopic data for 
these four Cepheids will be given in a forthcoming paper (Pedicelli et al.\ 2009, 
in preparation). We found accurate 
iron abundances for 77 out of the 173 Walraven Cepheids. Among them we selected 
the Cepheids with a reddening E(B-V)$\le$0.8 mag (66 objects). 

\section{Calibration of the  Metallicity--Index--Color relation}
\begin{figure}
   \centering
   \includegraphics[width=7.5cm]{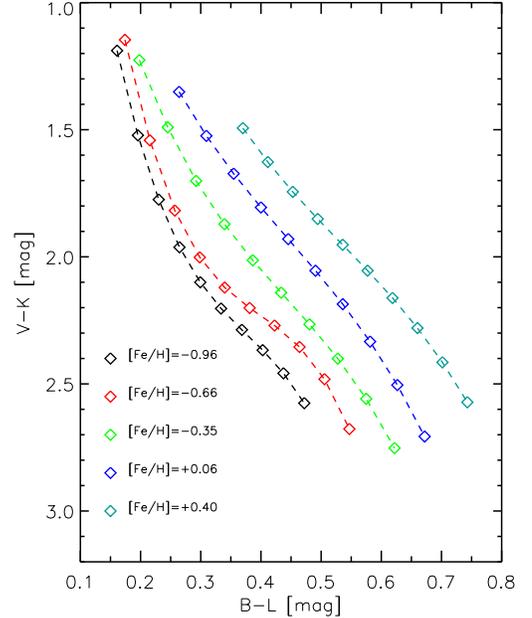}
   \caption {Predicted optical-NIR ($V-K$) -- Walraven ($B-L$) color-color plane 
for different chemical compositions (see labeled values). The dashed lines show the 
polynomial fits of the colors of helium burning phases (2nd and 3rd crossing) for 
different stellar masses and at fixed chemical composition. The diamonds display 
the uniformly sampled points adopted to determine the MIC relation.} \label{Fig2}
    \end{figure}

The spectroscopic data set does not allow us to provide a new empirical calibration 
of the Metallicity--Index--Color (MIC) relation for the Walraven bands, since 
the range in metallicity covered by these objects is quite limited 
(-0.4 $\le$$[Fe/H]$$\le$ 0.3 dex). To overcome this limitation we adopted 
several sets of evolutionary models characterized by different chemical 
compositions (Z=0.002, Y=0.248; Z=0.004, Y=0.251; Z=0.008, Y=0.256; 
Z=0.0198, Y=0.2734; Z=0.04, Y=0.303) and stellar masses ranging from 
5.0 to 10.0 $M_\odot$ (see Fig.\ref{Fig2}). The evolutionary tracks were 
computed by adopting a helium-to-metal enrichment ratio of $\Delta Y/\Delta Z\sim1.4$.       
For each evolutionary track we selected the helium burning phases 
(2nd and 3rd crossing) falling inside the predicted edges of the 
instability strip \citep{bo00,bo05}.  
We performed several sets of multilinear regressions between metal 
abundance and two independent Walraven, optical and NIR color indices. 
Eventually, we found the strongest sensitivity to metal abundance 
when using the Walraven $(B-L)$ and the optical-NIR $(V-K)$ color
(see Fig.\ref{Fig2}). 
The former color presents for warm stars a strong sensitivity to the 
metal content \citep{pelub07}, whereas the latter has a strong 
sensitivity to the effective temperature \citep{bo05}.  

In particular, we performed a polynomial fit of the colors of helium burning 
phases for the different stellar masses and at fixed chemical composition 
(see dashed lines in Fig.\ref{Fig2}). To improve the intrinsic accuracy of 
the MIC relation we uniformly sampled the polynomial fits (see diamonds in 
Fig.\ref{Fig2}) and adopted these points in the multi-linear regression.       
The theoretical MIC relations we derived can be parameterized as follows:  
\begin{equation}
  \begin{array}{c}
[Fe/H]_{phot} = -0.76[\pm 0.09] -0.55[\pm 0.45] \cdot (V-K) + \\ 
                +7.5[\pm 0.9] \cdot (B-L) -0.25[\pm 0.11] \cdot (V-K)^2 - \\
		-2.6 [\pm 0.9] \cdot (B-L)^2
   \end{array}
\end{equation}
where the symbols have the usual meaning and the number in parentheses are 
the uncertainties on the coefficients. 
By adopting this relation, we estimated the iron content for the entire 
Walraven data set. Interestingly enough, we found that the mean difference 
between photometric and spectroscopic iron abundances, for the 66 Cepheids 
for which the spectroscopic iron abundance is available, is as small as  
$\Delta[Fe/H]$ = -0.03 dex and the intrinsic dispersion is within 
photometric and spectroscopic uncertainties ($\sigma \sim $ 0.15 dex).

\section{Galactic metallicity gradient}
	In order to estimate the Galactic iron gradient the Galactocentric distances were 
	determined by assuming for the Sun a Galactocentric distance of 8.5 kpc 
	\citep{feastWhite97} and the classical formula for the Galactocentric 
        distances \citep{le07}. 

   \begin{figure}
   \centering
   \includegraphics[width=7.5cm]{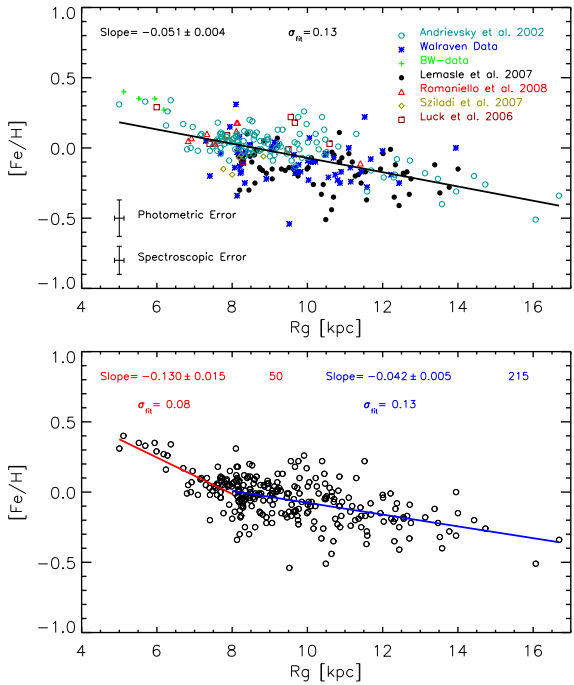}
   \caption {Top: Galactic abundance gradient based on the entire sample. 
   The different sets of spectroscopic data are marked with different symbols 
   and colors (see labels). The solid black lines shows the linear fit. The 
   slope and the standard deviation of the fit are also labeled. The error bars 
   in the left corner display the typical uncertainties on distances and on 
   photometric and spectroscopic abundances. Bottom: same as the top, but the 
   sample was split in Cepheids located either in the inner ($R_G < $8 kpc) or 
   in the outer disk. The red and blue line show the linear fit for the inner 
   (50 objects) and the outer (215 objects) disk, respectively. The slopes 
   and the standard deviations of the fits are also labeled.}
              \label{Fig3}
    \end{figure}

Together with the spectroscopic sample (207, see \S 2) we adopted 
58 new metallicity estimates based on the theoretical MIC relation and we 
ended up with a sample of 265 Cepheids, i.e.\ more than 50\% of the Galactic 
Cepheids currently known \citep{f95}. The top panel of Fig. \ref{Fig3} 
shows the radial distribution of the entire sample. In particular, the green crosses 
display the four metal-rich Cepheids for which we have new accurate abundances 
and new Baade-Wesselink distances, the photometric metallicities are marked with 
blue asterisks, while the spectroscopic abundances are marked with cyan open 
circles, \citet[][b, c]{and02a}; black filled circles, \citet{le08}; 
red triangles, \citet{ro08}; yellow diamonds, \citet{szi07}; 
purple squares, \citet{lu06}. 
A glance at the data plotted in this panel shows that spectroscopic and photometric 
metallicities present similar trends. A linear fit over the entire sample gives an 
iron abundance gradient with a slope of -0.051$\pm$0.004 dex $kpc^{-1}$. This 
estimate agrees quite well with the most recent literature values based either 
on Cepheids \citep[][-0.052$\pm$0.003]{le08} or on open clusters 
\citep[][-0.06$\pm$0.01]{fri02}. However, the linear fit plotted in the top panel 
of Fig. \ref{Fig3} is  systematically more metal-poor than observed iron abundances 
in the innermost disk regions ($R_G$$\lesssim$7 kpc). By using open clusters 
\citet{twa97} suggested the occurrence of a sharp discontinuity in the metallicity 
gradient at $R_G\approx$ 10 kpc, while \citet{carr07} found evidence of a flattening 
in the outer disk. The occurrence of a discontinuity in the outer disk was also 
supported by \citet{lu03} and by \citet{and02c, and04} using Cepheids. On the other hand,  
the occurrence of a steepening of the metallicity gradient in the inner disk 
was suggested by \citet{and02c} using Cepheids. 
Data plotted in the top panel of Fig. \ref{Fig3} do not support the presence of 
a discontinuity for $R_G\approx$ 10-12 kpc. This finding is in agreement with 
the results of \citet{le08}. Instead, current data show a smooth and steady 
increase in the slope for $R_G<$ 8 kpc. Therefore, we split the Cepheids 
according to their Galactocentric distance in an inner ($R_G<$ 8 kpc, 50) 
and in an outer (215) subsample.   
The bottom panel of Fig. \ref{Fig3} shows that the iron gradient in the inner 
disk is more than a factor of three steeper (-0.130$\pm$0.015 dex $kpc^{-1}$, 
$\sigma_{fit}$=0.08 dex, red line) than the slope in the outer disk 
(-0.042$\pm$0.004 dex $kpc^{-1}$, $\sigma_{fit}$=0.13 dex, blue line). 
In passing, we also note that the current linear fits indicate an iron 
abundance of $\approx$0.3 dex in the innermost disk regions and of 
$\approx$-0.3 dex in the outermost disk regions. The difference ($\approx$0.6 dex) 
and the flattening of the gradient in the outer disk are still hampered 
by the limited number of Cepheids in this crucial region. 

\begin{figure}
   \centering
	   \includegraphics[width=7.5cm]{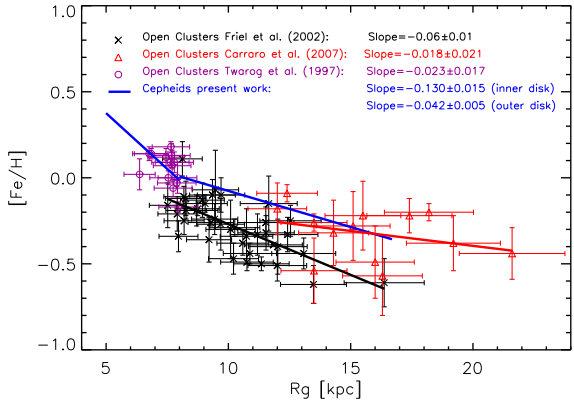}
   \caption{Comparison between iron abundance gradients based either on 
open clusters (\citet{twa97}, purple dots; \citet{fri02}, black dots; 
\citet{carr07}, red dots) or on Cepheids (blue lines). Note that for 
the sample collected by \citet{twa97} we only plotted the open clusters 
with $R_G < 8$ kpc. The error bars display individual uncertainties on 
iron abundances and on distances.}
	      \label{Fig4}
  \end{figure}

Fig. \ref{Fig4} shows the comparison between the current iron 
abundance gradients and similar gradients based on open clusters provided 
by  \citet[][purple dots]{twa97}, by \citet[][black dots]{fri02} and 
by \citet[][red dots]{carr07}. Note that to avoid the crowding between 
the different samples of open clusters, for the Twarog et al. sample 
we only plotted the clusters with $R_G\le$8 kpc.   
Data plotted in this figure show that the Cepheid slope for the outer 
disk agrees, within the errors, quite well with the slope estimated by 
\citet{fri02}, while it appears steeper than the slope estimated by 
\citet{carr07}. The difference in slope is due to a difference 
in the Galactocentric distribution of the adopted tracers. The flattening
in the slope by \citet{carr07} is mainly based on four open clusters with 
$R_G>$ 17 kpc. Unfortunately, we still lack accurate iron abundances for 
Cepheids located at large Galactocentric distances. The comparison also 
indicates that in the inner disk the slope based on Cepheids is significantly 
steeper (-0.13 vs -0.018 dex kpc$^{-1}$) than the slope based on open cluster 
collected by \citet{twa97}. The outcome is the same if we only account for 
open clusters with $R_G\le$8 kpc, i.e. -0.13 vs -0.073 dex kpc$^{-1}$.  
We cannot reach a firm conclusion concerning the difference between the 
two different slopes, since the Twarog's  sample includes only one 
cluster with  $R_G\le$6.5 kpc.  
Finally, we mention that the difference, at fixed Galactocentric distance, 
between Cepheid and open cluster abundances could be partially due 
either to a difference in the metallicity scale or to an age effect. 
The first problem might be addressed using Galactic cluster Cepheids 
\citep{fryCarney97}, but we still lack firm quantitative constraints.  
The open clusters in the Friel's sample cover a broad 
age range ($\sim$0.8 -- $\sim$9 Gyr), while Cepheids are young He-burning 
stars and their ages using the Period-Age relation provided by \citet{bo05} 
range from $\approx$ 10 to $\approx$ 130 Myr. Empirical evidence 
\citep{fri02} and theoretical predictions \citep{por99, por00} indicate that 
the iron gradient flattens as a function of decreasing age. However, firm 
quantitative constraints on this effect require homogeneous estimates 
of distances and iron abundances for the two different tracers.     

\section{Discussion and conclusions}

The data plotted in Fig.\ref{Fig3} show a significant 
spread in metallicity at fixed Galactocentric distance. A similar evidence 
was brought forward by \citet{lu06} and confirmed by \citet{le08}. The 
former authors, using spectroscopic iron abundances for 205 Cepheids, found 
evidence of abundance inhomogeneities across the disk. In particular, they 
found a group of stars approaching supersolar iron abundances (+0.2 dex) 
for $R_G\approx$~9.5-10 kpc. The latter authors, using spectroscopic iron 
abundances for 178 Cepheids, found that the spread in metallicity for 
$R_G\approx$~10-12 kpc was strongly correlated with the Galactic longitude 
(see their Figures 5 and 10). 
In order to analyze quantitatively the available metallicity data for 
Cepheids, we performed several tests by selecting the 
Cepheids either according to their age (period cuts) or to the height 
above the Galactic plane. The spread in iron content was minimally affected 
by these selections. Therefore, we  projected the positions of our 265 
Cepheids onto the Galactic plane. The data plotted in Fig.\ref{Fig5} 
show that our sample covers the four Galactic quadrants quite well.
The relative fractions per quadrant are: 
21.1$\pm$0.8\% (Q1), 32.1$\pm$0.7\% (Q2), 26.4$\pm$0.7\% (Q3) 
and 21.4$\pm$0.8\% (Q4). The errors on the relative fractions 
only account for Poisson uncertainties.  
Moreover, to constrain their radial distribution 
across the Galactic disk we split the Cepheids in a metal-poor 
(MP, $[Fe/H] < -0.02$) and in a metal-rich (MR, $[Fe/H] \ge -0.02$) 
subsample with similar number of objects (134 vs 131). The blue (MP) 
and the red (MR) dots plotted in Fig.\ref{Fig5} disclose several 
interesting features:

   \begin{figure}
   \centering
   \includegraphics[width=8.cm]{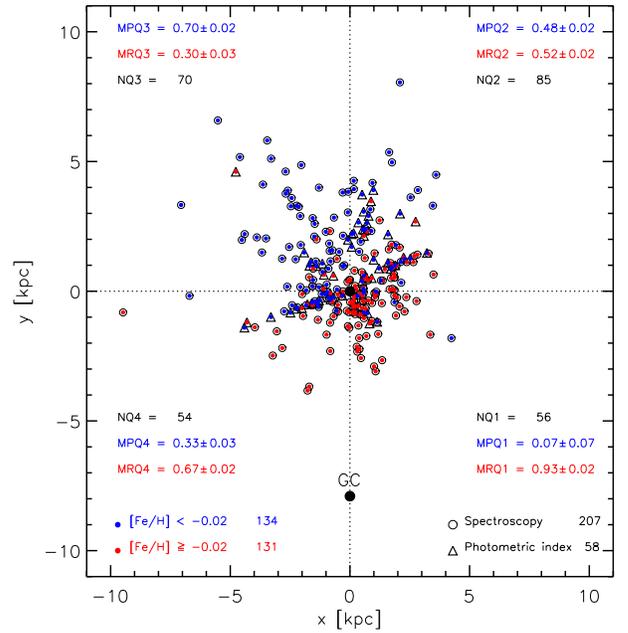}
   \caption{Cepheid distribution projected onto the Galactic plane. 
   Open circles mark the 207 Cepheids with spectroscopic iron abundances, 
   while the triangles mark the 58 Cepheids with photometric metallicities. 
   The blue and the red dots display metal-poor (MP, $[Fe/H]<$-0.02, 134) 
   and metal-rich (MR, $[Fe/H]$$\ge$-0.02 131) Cepheids. The total number  
   of Cepheids and the relative fractions of MP and MR Cepheids per quadrant 
   are also labeled. The errors on the relative fractions only account for 
   Poisson uncertainties. The black circle marks the Galaxy center (GC).}
            \label{Fig5}
    \end{figure}

\begin{figure}
   \centering
   \includegraphics[width=8.cm]{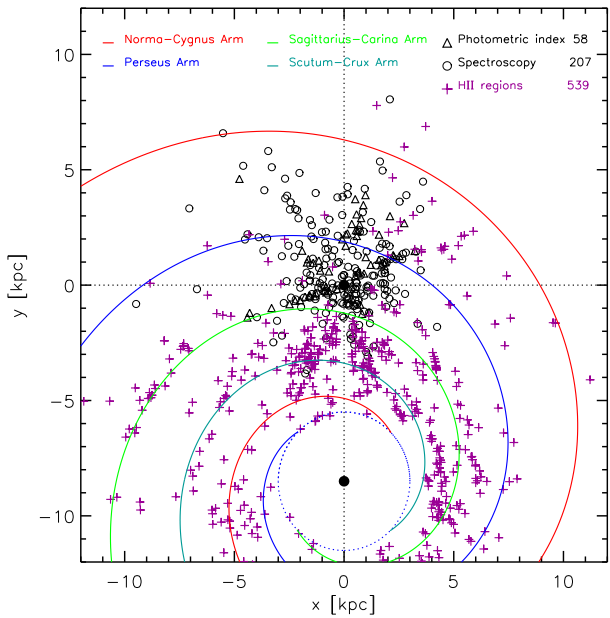}
   \caption{Cepheid distribution projected onto the Galactic plane. 
   The symbols are the same as in Fig. \ref{Fig5} and the purple 
   pluses mark $HII$ regions collected by \citet{pal04}. The four 
   spiral arms according to the Galactic model of \citet[][]{val05} 
   are plotted with different colors (see labels).}
              \label{Fig6}
    \end{figure}

\begin{itemize}
\item{\em Metallicity Distribution --} The metallicity distribution 
in the four quadrants does not show a smooth trend when moving from the 
innermost to the outermost disk regions.  We found that in the first 
quadrant the bulk of Cepheids is MR ($\sim93\pm2$\%), 
while in the fourth quadrant the relative fractions are 
$67\pm2$\% (MR) and $33\pm3$\% (MP), respectively. Note that the 
errors on the relative fractions only account for Poisson uncertainty.   
The asymmetry in the metallicity distribution is also present in the 
top quadrants. The second quadrant shows, within the errors, similar 
fractions of MP ($48\pm2$\%) and MR ($52\pm2$\%) Cepheids, 
while in the third quadrant a significant fraction of Cepheids are MP 
($70\pm2$\%) and less than one third are MR ($30\pm3$\%). 
Our findings do not indicate a strong asymmetry in the metallicity 
distribution between bottom and top quadrants, but an asymmetry 
between the first and the third quadrant. The difference is at more  
than 8$\sigma$ level for the MP and more than 15$\sigma$ level for 
the MR subsample.     
   
\item{\em Local Inhomogeneities --} Abundance inhomogeneities occur not 
only at level of quadrants, but also on smaller spatial scales. 
The iron abundance of the Cepheids belonging to the two 
overdensities located in the second (X$\sim$2, Y$\sim$1 kpc) and in the 
fourth quadrant (X$\sim$-1.5, Y$\sim$-0.5 kpc) ranges 
from $\sim$-0.20 to $\sim$0.25 dex and from $\sim$-0.30 to $\sim$0.20 dex, 
respectively. This means that in these regions the Cepheids approximately 
cover the same range in metallicity covered by the global gradient.   

\item{\em Spatial Distribution --} The Cepheid spatial distribution 
across the disk is reminiscent of Galactic spiral arms. After the seminal 
investigations by \citet{kraSchm63} it has been recently suggested 
by \citet{lu06} that the overdensity they detected in the second 
quadrant (at $l\approx 120^{\circ}$ and at about 3 kpc from the Sun) is 
located in the same region of the Perseus arm. 
To analyze the correlation between the spatial distribution of Cepheids  
and spiral arms across the Galactic plane we plotted in Fig.\ref{Fig6} 
Cepheids together with $HII$ regions collected by \citet{pal04}. 
Note that kinematic distances of $HII$ regions might 
be affected by systematic uncertainties \citep{rei09b}. Moreover, 
to help the identification with the spiral arms we also plotted the 
four arms according to the simple Galactic model of \citet{val05}. 
Note that no general consensus has been reached yet concerning the 
number of spiral arms, the pitch angles and the number of warps 
(Reid et al.\ 2009, and references therein). The arms plotted 
in Fig.\ref{Fig6} are only aimed at a preliminary qualitative correlation.        
The data plotted in this figure further support the association of the 
overdensity located at X$\sim$2, Y$\sim$1 kpc (second quadrant) with 
the Perseus arm. Moreover, the overdensity in the fourth quadrant 
(X$\sim$-1.5, Y$\sim$-0.5 kpc) covers more than 2 kpc and it appears 
to be associated with a group of $HII$ regions and with the 
Sagittarius-Carina arm. The small overdensity located in the first quadrant 
(X$\sim$0.5, Y$\sim$-1.0 kpc) is associated with an overdensity of 
$HII$ regions located between the Sagittarius-Carina and the Scutum-Crux 
arm. In the third quadrant (X$\sim$-2.5, Y$\sim$3.5 kpc) and in the 
first quadrant are also present two overdensities, but the number of 
$HII$ regions is either limited or missing.   
\end{itemize}
    
The above findings support recent findings by \citet{lu06} 
and by \citet{le07} concerning the occurrence of chemical 
inhomogeneities across the Galactic quadrants. Moreover and even more 
importantly, Cepheid abundances indicate that these inhomogeneities are also 
present at smaller spatial scales. This means that the global abundance 
gradients should be cautiously treated, since empirical evidence indicates  
a chemical enrichment with a clumpy distribution across the Galactic disk.    

The use of Classical Cepheids not only to trace the chemical enrichment 
across the disk, but also to constrain together with open clusters, 
$HII$ regions and star forming regions its chemical tagging 
\citep{desi07} appears particularly promising. The use of homogeneous 
metallicity scales for field and cluster stars \citep{yo05, yo06, bra09} 
and homologous distance scales \citep{fou07,groe08} 
should also be explored further. The comparison of metallicity gradients based 
on different stellar tracers and robust constraints on the age dependence of 
the metallicity gradient rely on these two relevant requirements.    
 
\begin{acknowledgements}
It is a real pleasure to thank S. Paladini for sending us the data on $HII$ 
regions in electronic form. We also thank C. Chiosi and P. Prada Moroni 
for many useful discussions on Galactic chemical evolution and on Galactic 
models. One of us (SP) thanks ESO  for the PhD studentship.   
This publication makes use of data from the Two Micron All Sky Survey, which 
is a joint project of the University of Massachusetts and the Infrared Processing 
and Analysis Center/California Institute of Technology, funded by the National 
Aeronautics and Space Administration and the National Science Foundation.
We also thank the ESO Science Archive for their prompt support and the anonymous 
referee for his/her positive opinion concerning the content of this investigation.
\end{acknowledgements}

\bibliographystyle{aa}
\bibliography{mybib}

\end{document}